\documentclass[aps,epsfig,floats,twocolumn,amssymb,amsmath,showpacs]{revtex4}
\usepackage{graphicx}
\begin{document}
\title{Aggregation kinetics of stiff polyelectrolytes in the presence of multivalent salt}

\author{Hossein Fazli}
\affiliation{Institute for Advanced Studies in Basic Sciences,
Zanjan 45195-1159, Iran}

\author{Ramin Golestanian}
\email{r.golestanian@sheffield.ac.uk} \affiliation{Department of
Physics and Astronomy, University of Sheffield, Sheffield S3 7RH,
UK}

\date{\today}

\begin{abstract}
Using molecular dynamics simulations, the kinetics of bundle
formation for stiff polyelectrolytes such as actin is studied in the
solution of multivalent salt. The dominant kinetic mode of
aggregation is found to be the case of one end of one rod meeting
others at right angle due to electrostatic interactions. The kinetic
pathway to bundle formation involves a hierarchical structure of
small clusters forming initially and then feeding into larger
clusters, which is reminiscent of the flocculation dynamics of
colloids. For the first few cluster sizes, the Smoluchowski formula
for the time evolution of the cluster size gives a reasonable
account for the results of our simulation without a single fitting
parameter. The description using Smoluchowski formula provides
evidence for the aggregation time scale to be controlled by
diffusion, with no appreciable energy barrier to overcome.
\end{abstract}

\maketitle

\section{Introduction}  \label{sec:intro}

Highly charged polyelectrolytes (PEs) are known to attract each
other due to electrostatic correlations brought about by multivalent
counterions (ions of opposite charge) \cite{reviews}. A ubiquitous
phenomenon arising from these correlations is the formation of
collapsed bundles of stiff PEs \cite{Bundle,Angelini}, which is
believed to play a significant role in biological processes such as
cell scaffolding dynamics \cite{cyto} and motility \cite{mogilner}.
While our current theoretical understanding of the process of PE
bundle formation predicts a macroscopic phase separation, i.e. an
infinitely large bundle \cite{equ-inf}, experiments always find
finite-sized bundles \cite{Bundle,Angelini}. To explain this, it has
been suggested that the theoretically expected phase separation may
be hindered by kinetic barriers \cite{HaLiu}, steric effects
\cite{Henle}, or frustration of the local structure with energy
penalty \cite{Angelini,grason}. The phenomenon of bundle formation
has also been studied using computer simulation, which indicated a
tendency towards a well defined finite size \cite{simu,Holm}.
Recently, an extensive study on the thermodynamic properties of a
condensed bundle with multivalent counterions has been carried out,
which shows that finite bundles are stable for an intermediate range
of electrostatic couplings whereas at strong enough coupling the
bundle could be macroscopic \cite{Sayar}. Multivalent counterions
can also induce the structural collapse of single semiflexible
polyelectrolytes \cite{collapse} and highly charged polyelectrolyte
brushes \cite{brush}.

We can understand more about this phenomenon by studying the
kinetics of the bundle formation. The angle dependent interaction
between two rods has been recently studied \cite{Lee} and it has
been shown that the preferred relative orientation of the two rods
has a non-trivial connection with whether the overall interaction is
attractive or repulsive. One would then like to know how this
complicated angle-dependent interaction affects the fate of the
filament bundle in the course of the aggregation kinetics.

Another question could be the dominant kinetic mode of aggregation.
A Brownian dynamics simulation using a model short-ranged attraction
between two rods has shown that the rods are most likely to meet in
a cross configuration, presumably against a barrier as studied in
Ref. \cite{HaLiu}, and then rotate and slide to adjust into a
parallel pair \cite{Yu}. Another pathway with a lower barrier has
been suggested in which a rod will slide parallel to an existing
bundle \cite{Shkl}. While these descriptions are typically in terms
of pairs of rods or an effective interaction of a single filament
with an already formed bundle, it is not clear {\em a priori} that
they are feasible in such a highly correlated system. In particular,
the orientational dependence of the many-body interaction between
highly charged polymers at close proximity is highly complex and can
lead to nontrivial effects, such as spontaneous formation of chiral
structures \cite{fazli}. Recently, an experiment has been performed
to probe the kinetics of bundle formation by using two different
fluorescently labeled actin filaments, which suggests that actin
bundles dynamically exchange filaments with the solution
\cite{GH-Lai}. In light of the new experimental insight into the
system, it is important to set up a more systematic study of the
kinetics of the aggregation, so that the effect of the nontrivial
many-body correlations during the course of the aggregation can be
better understood.

\begin{figure}[t]
\includegraphics[width=.99\columnwidth]{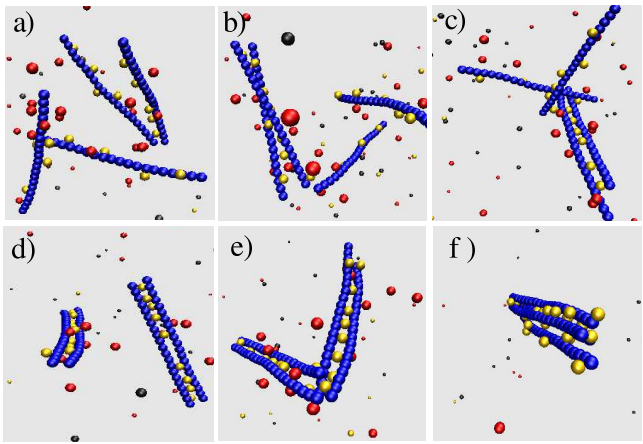}
\caption{(color online). Time lapse snapshots of the system with
$N_p=4$ and $c_{3:1}=1$. $+3$ salt ions are shown by golden (light)
spheres, $+1$ counterions by red (dark), and $-1$ salt ions by gray
spheres. It can be observed that when two filaments or two bundles
meet each other the dominant crossing angle is $90$ degrees. In the
final configuration all the PEs form a single
bundle.}\label{fig:snapshot1}
\end{figure}

\begin{figure}[t]
\includegraphics[width=.99\columnwidth]{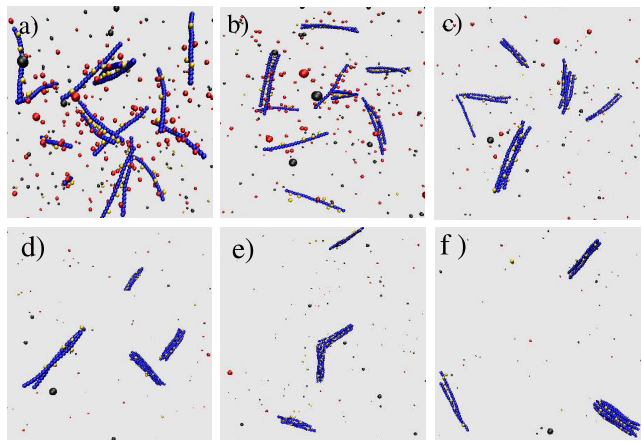}
\caption{(color online). Time lapse snapshots of the system with
$N_p=16$ and $c_{3:1}=1$. The largest bundle contains 11 PE rods
(see part f of the figure) and two smaller bundles containing 2 and
3 PEs finally join up and make a bundle of 5 rods (final
configuration is not shown in this figure).}\label{fig:snapshot2}
\end{figure}

Here, we study the kinetics of bundle formation in the bulk solution
of stiff polyelectrolytes in the presence of multivalent salt using
molecular dynamics (MD) simulations. We find that the PEs undergo an
aggregation process with doublets, triplets, etc forming and
subsequently feeding into larger clusters. We observe that the
initial stage of the kinetics leads to formation of PE bundles that
have a clear size selection, up to 10-11 filaments for our choice of
parameters. These bundles take up all the smaller clusters, and are
relatively much more long-lived than the smaller ones, while larger
clusters do not seem to appear even when there are a number of these
long-lived filament bundles available in the solution for possible
aggregation. We find that the time-dependent size distribution of
the aggregates follows a Smoluchowski flocculation kinetics. For the
first three cluster sizes, namely single filaments, doublets, and
triplets, the time evolution of the number of such clusters shows a
reasonable quantitative agreement with the Smoluchowski formula when
the energy barrier is set to zero. This result shows that the
many-body energy barrier for the formation of these clusters is
relatively small, and that the time scale for the evolution is set
by the diffusion of filaments in the course of the aggregation
process. We also monitor the kinetics of bundle growth and find that
the dominant mode is due to one end of a filament/bundle meeting
another filament/bundle either in the middle mostly at right angle
(in the form of ``$\dashv$'') or at one end (in the form of
``$\left\rangle\right.$'') before rotating and sliding into a
parallel packing (see Figs. \ref{fig:snapshot1} and
\ref{fig:snapshot2}). We show that these modes can be understood
from energetic considerations by calculating the angle dependence of
the potential of mean force between two rods.

The rest of the paper is organized as follows. Section \ref{sec:sim}
introduces the model and details of the simulation technique used in
this work. The main results from the simulation are presented in
Sec. \ref{sec:result}, while Sec. \ref{sec:eff} is dedicated to the
potential of mean force of two rods at various angles and
separations. Finally, Sec. \ref{sec:discuss} closes the paper with
some discussions and remarks.

\section{The Simulation}    \label{sec:sim}

In our simulations of the bulk solution, which are performed with
the MD simulation package ESPResSo v.1.8 \cite{Espresso}, $N_p$
stiff PEs are considered each composed of $N_m=21$ spherical charged
monomers of charge $-e$ (electronic charge) and diameter $\sigma$
which is introduced via the Lennard-Jones potential (see below).
Monomers of each PE are bonded to each other with separation between
them being fixed at $1.1 \sigma$ via a finite extensible nonlinear
elastic (FENE) potential \cite{Grest} and the bending rigidity of PE
chains is modeled with a bond angle potential
$U_{\phi}=k_{\phi}(1-\cos{\phi})$ with $k_{\phi}=400 \;k_{\rm B}T$
in which $\phi$ is the angle between two successive bond vectors
along the PE chain. We use $N_c=N_p \times N_m$ monovalent
counterions with charge $+e$ to neutralize the PEs. We also consider
trivalent salt with $N^+_{s}$ positive ions with charge $+3 e$ and
$N^-_{s}=3 N^+_{s}$ negative ions with charge $-e$. In addition to
long-ranged Coulomb interaction we include short-ranged
Lennard-Jones repulsion between particles, which introduces an
energy scale $\epsilon$ into the system. MD time step in our
simulations is $\tau=0.01\tau_0$, in which
$\tau_0=\sqrt{m\sigma^2/\epsilon}$ is the MD time scale and $m$ is
the mass of the particles.

\begin{figure*}[t]
\includegraphics[width=1.9\columnwidth]{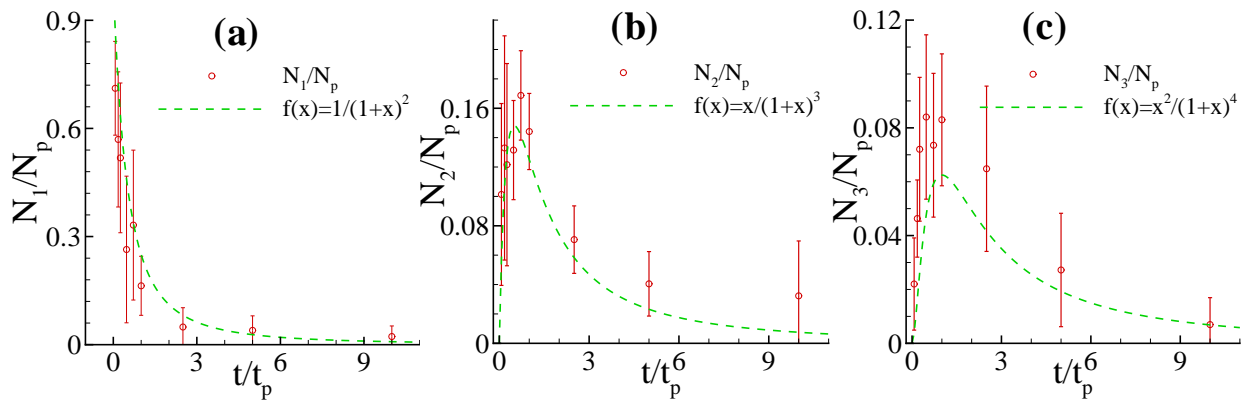}
\caption{(color online). $N_k/N_p$ versus rescaled time $t/t_p$ for
(a) $k=1$, (b) $k=2$, and (c) $k=3$, with $W=1$. Data and error bars
are obtained from averaging over sets of data such as those
presented in Table \ref{tab:table1} corresponding to various
simulations with identical and differing values of $N_p$. }
\label{fig:n1_n2_n3}
\end{figure*}

The temperature is fixed at $k_{\rm B} T=1.2\epsilon$ using a
Langevin thermostat. We use the particle-particle particle-mesh
(PPPM) method to apply periodic boundary conditions for long-ranged
Coulomb interaction in the system. The strength of the electrostatic
interaction energy relative to the thermal energy can be quantified
using the Bjerrum length $\ell_{\rm B}=\frac{e^2}{\varepsilon k_{\rm
B}T}$, where $\varepsilon$ is the dielectric constant of the solvent
and in our simulations we have used it to fix the value for $\sigma$
via $\ell_{\rm B}=3.2\sigma$. Following Ref. \cite{Lee}, we define
the salt concentration as $c_{3:1}=N^-_{s}/N_c$, and use values in
the range of $c_{3:1}=0.5-1.2$. The Debye length $\lambda_{\rm
D}=6\sigma$ for $c_{3:1}=1$.

In the beginning of the simulations, we fix the PE rods in space
parallel to each other with their centers arranged on a square
lattice of spacing in the range of $10\sigma-15\sigma$ in the middle
of the simulation box and all of the ions are free to move for
100,000 MD time steps. We then release the PEs and after
equilibration of the system we study the bundle formation kinetics
for 8,000,000 MD steps.

\section{Results} \label{sec:result}

We do simulations for different values of the number of PEs $(N_p=4,
9, 16, 25, 64)$ and find that in simulations with $4$ and $9$ PE
rods, in the final configuration of the system a single bundle of
parallel PEs forms containing all of them. Figure
\ref{fig:snapshot1} shows snapshots of the system with $N_p=4$ PE
rods at different times. It can be seen that PEs that are going to
be added to a bundle of parallel PEs, first meet the bundle at right
angle and then the crossing angle vanishes.

Simulations with larger $N_p$ show the same kinetic process (see
Fig. \ref{fig:snapshot2}) as well as an aggregation with small
clusters forming initially and then feeding into larger clusters.
Table \ref{tab:table1} shows the numbers of PE bundles of different
sizes at different times for the simulation with $N_p=25$. In this
table, $N_k$ is the number of bundles containing $k$ PEs and for a
system with $N_p$ PEs it is subject to the normalization $\sum_{k}{k
N_k}=N_p$. The largest bundle at long times contains $11$ PEs in all
the simulations. In the case of $N_p=16$, two bundles containing $5$
and $11$ rods remain at long times with no affinity towards each
other when we ran the simulation for longer times. Simulations with
larger numbers of PEs, i.e. $N_p=25$ (see Table \ref{tab:table1})
and $64$ confirm that the growth of the bundles tend to be cut off,
or slowed down beyond the time span of our simulation, when there
are 10-11 filaments in the bundle. While at the end of the
simulation there are more bundles as the number of rods increases,
the largest bundle at long times remains to be made of maximum 11
rods.

\begin{table}[b]
\caption{\label{tab:table1}Distribution of PE bundle sizes for
different values of MD time for a system with $N_p=25$ and
$c_{3:1}=1$.}
\begin{ruledtabular}
\begin{tabular}{cccccccc}
${\rm time}(\times 2000\tau)$ &$N_1$ & $N_2$
&$N_3$ &$N_4$ &$N_5$ &$N_6$ &$N_{10}$\\
\hline 10& 18 & 2 & 1 &0
& 0 & 0 & 0 \\
35& 14 & 4 & 1 &0
& 0 & 0 & 0 \\
42& 13 & 3 & 2 &0
& 0& 0 & 0 \\
92& 10 & 3 & 3 &0
& 0 & 0 & 0 \\
144& 9 & 3 & 2 &1
& 0 & 0 & 0 \\
158& 8 & 3 & 1 &2
& 0 & 0 & 0 \\
270& 7 & 2 & 0 &2
& 0 & 1 & 0 \\
350& 5 & 3 & 0 &2
& 0 & 1 & 0 \\
400& 4 & 3 & 0 &1
& 1 & 1 &0 \\
450& 4 & 3 & 0 &0
& 1 & 0 & 1 \\
750& 3& 2 & 1 &0 &1 &0 &1 \\
1500& 2 & 2 & 0 &1
& 1 & 0 & 1 \\
2000& 2& 0 &0 &2 &1 &0 &1 \\
2700& 1& 0& 0& 1& 0& 0& 2 \\
2900& 0& 0& 0& 0& 1& 0& 2 \\
4000& 0& 0& 0& 0& 1& 0& 2 \\
\end{tabular}
\end{ruledtabular}
\end{table}

The evolution of the clusters and their distribution---as
exemplified in Table \ref{tab:table1} for 25 rods---resembles the
aggregation kinetics of colloidal particles \cite{colloid}. For the
case of spherical colloidal particles with short-ranged
interactions, Smoluchowski suggested an (approximate) expression for
the number of clusters of size $k$ in time $t$ as \cite{colloid}
\begin{equation}
N_k(t)=\frac{N_p (t/t_p)^{k-1}}{(1+t/t_p)^{k+1}}. \label{smol-eq}
\end{equation}
In this equation, the characteristic time is defined as
\begin{equation}
t_p=\frac{\eta W}{n_0 k_{\rm B}T},\label{tp-eq}
\end{equation}
where $\eta$ is the viscosity of water, $n_0$ is the initial number
density of the particles, and $W \simeq e^{E_a/k_{\rm B} T}$ is an
activation factor (up to a numerical coefficient). The values of
$N_k/N_p$ for $k=1$, $2$ and $3$ obtained from our simulations are
plotted in Fig. \ref{fig:n1_n2_n3} as a function of ${t}/{t_p}$ with
$W=1$. In these plots, we have taken $\eta\simeq 2.4
\sqrt{m\epsilon}/\sigma$ (from Ref. \cite{Dunweg}), $k_{\rm B}
T=1.2\epsilon$, $n_0=0.001\;\sigma^{-3}$, and
$\tau=0.01\tau_0=0.01\sqrt{m\sigma^2/\epsilon}$, which yields
$t_p=2\times 10^5\;\tau W$. The results are compared in Fig.
\ref{fig:n1_n2_n3} with the Smoluchowski formula of Eq.
(\ref{smol-eq}) and an agreement in the range of the error bars can
be seen. Naturally, the fact that the clusters do not exceed the
maximum size of 10-11 means that the analogy is limited to the
smaller sizes. Nevertheless, it is remarkable that the same value
for $t_p$ gives a good agreement for $N_1/N_p$, $N_2/N_p$ and
$N_3/N_p$, without even a single fitting parameter being used.

The fact that $W=1$ seems to provide a reasonable agreement suggests
that the PE rods do not experience substantial energy barriers in
their ``optimal'' kinetic paths in the course of the aggregation,
although an exact knowledge of the numerical prefactor for $W$ in
the rod geometry is required for a precise determination of the
``activation energy'' $E_a$ (which may even turn out to be
negative). Note that the low optimal energy barrier does not mean
that energetics does not play a role in the kinetics, as the
dominant kinetic mode is clearly a result of strong electrostatic
interactions.

The problem of quantifying the kinetic barrier of the aggregation
process is a complicated one, as it is not clear how one can
approach it better than just looking at pair-potentials. Here we
propose that monitoring the time evolution of the cluster sizes
could be a good alternative for capturing the many-body essence of
the aggregation process. The typical time scale $t_p$ that is
involved in the evolution of the cluster sizes [see Eq.
(\ref{tp-eq})] has a diffusion-controlled component $\eta/(n_0
k_{\rm B}T)\sim \ell^3/(L D)$, where $\ell=n_0^{-1/3}$ is the
initial average distance between the filaments in the solution, $L$
is the length of the filaments, and $D$ is a filament diffusion
coefficient. The second component of $t_p$ comes from the many-body
energetics of the system. A comparison of the simulation results for
the cluster sizes with the Smoluchowski formulas in Fig.
\ref{fig:n1_n2_n3} suggests that the diffusion part is dominant
during the early stages of the aggregation until the finite-sized
bundles are formed. The fact that the Smoluchowski plots for zero
energy barrier are close to the simulation data points, shows that
the energy barriers have to be small. We chose to plot the zero
energy barrier form instead of trying to fit the data to the
Smoluchowski formula and deduce an energy barrier, because we do not
know the right prefactors for the different geometry of rods (rather
than spheres in the original Smoluchowski solution). However, these
prefactors must not differ too much from unity and therefore the
argument that the barrier must be small at this stage will be
justified from the agreement in Fig. \ref{fig:n1_n2_n3}.

\begin{figure}[t]
\includegraphics[width=.9\columnwidth]{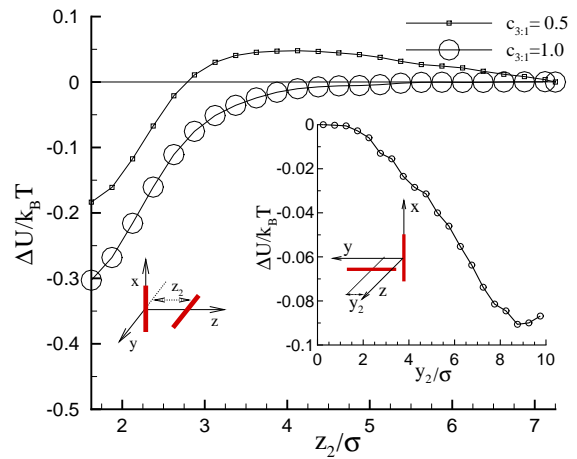}
\caption{(color online). Potential of mean force per monomer between
two perpendicular rods as a function of the separation between their
centers $z_2$. The size of the circles on $c_{3:1}=1.0$ curve
corresponds to the error bars. Inset: potential of mean force
between two rods which are perpendicular as a function of $y_2$ with
$c_{3:1}=0.7$.} \label{fig:deltaU_z2_y2}
\end{figure}

\section{Angular Dependence of the Potential of Mean Force}
\label{sec:eff}

To understand the dominant kinetic mode of PE aggregation we study
the different configurations of two rods and keep them fixed during
simulation while allowing the counter-ions and salt ions to move
freely. In these simulations, we calculate the average force on each
monomer and obtain the total force and torque for each PE rod. By
integrating the average force (torque) over different separations
(orientation angles), we obtain the potential of mean force over a
displacement (rotation about an axis) \cite{Lee}. Let us denote the
coordinates of the centers of PEs 1 and 2 as $(x_1,y_1,z_1)$ and
$(x_2,y_2,z_2)$ respectively (see the schematic configuration of the
rods in Fig. \ref{fig:deltaU_z2_y2}).

We first show that in the range of salt concentration we use in our
simulations, there is an attractive interaction between two PE rods
which are perpendicular to each other. Assume that rod 1 is fixed on
the $x$ axis with its center at the origin, and rod 2 is parallel to
the $y$ axis with the coordinates of its center being $(0,0,z_2)$.
Figure \ref{fig:deltaU_z2_y2} shows the potential of mean force,
which is calculated as the reversible work to bring rod 2 from
$z_2=7.5\sigma$ to closer separations as a function of $z_2$, for
two values of salt concentration $c_{3:1}=1$ and $0.5$. It can be
seen that there is an attractive interaction between two
perpendicular PEs. We also calculate the potential of mean force
between two perpendicular rods when $z_2$ is kept fixed at
$z_2=1.5\sigma$, and $y_2$ is changed (see the schematic
configuration in the inset of Fig. \ref{fig:deltaU_z2_y2}) for
$c_{3:1}=0.7$. We find that the potential of mean force decreases
with increasing $y_2$, which shows that when the two rods are
perpendicular at close separations, the best configuration is when
one end of one rod touches the other at right angle (although the
potential of mean force is only slightly lower).

\begin{figure}[t]
\includegraphics[width=.9\columnwidth]{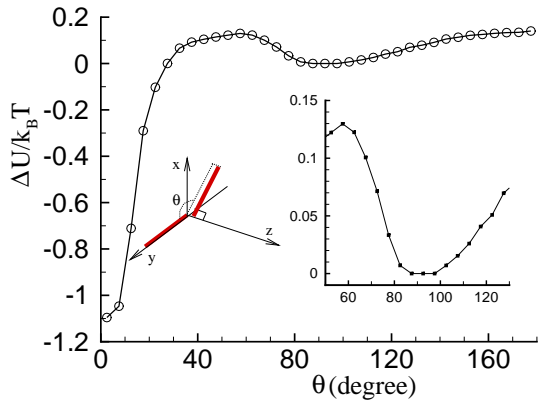}
\caption{(color online). The potential of mean force per monomer as
a function of $\theta$ for $c_{3:1}=0.7$. A local minimum can be
seen around $\theta=90^\circ$, which is also shown as the inset of
the figure. The size of the circles in the main figure shows the
error bar.}\label{fig:deltaU_theta}
\end{figure}

We also fix rod 1 on the $x$ axis with its end at the origin and fix
one end of rod 2 at $(0,0,1.5\sigma)$ (see Fig.
\ref{fig:deltaU_theta}), and calculate the angular dependence of the
potential of mean force by integrating the torque with respect to
the rotation angle $\theta$ (around $z$ axis). In Fig.
\ref{fig:deltaU_theta}, the orientational dependence of the
potential of mean force of the two rods is shown relative to the
$\theta=90^\circ$ configuration. In this figure, we can see that
although the global minimum of the potential of mean force
corresponds to parallel configuration of the rods, a local minimum
exists around $\theta=90^\circ$, which is shown in detail in the
inset of the figure. The attractive interaction between two parallel
PEs in the presence of multivalent salt appears only at short
separations \cite{Lee}. This means that for two rods at larger
separations, it is more likely that they attract each other when
they have larger relative angles, and one of the ends is meeting the
other PE in the solution. Moreover, Fig. \ref{fig:deltaU_theta}
shows that although the local minimum is shallow compared to the
parallel-configuration minimum, it covers a relatively wider range
of $\theta$, and thus crossing at ``nearly'' right angle has a
considerably high probability.
For obtaining each point of Figs. \ref{fig:deltaU_z2_y2} and
\ref{fig:deltaU_theta}, we have used $1.5\times 10^5$ MD steps for
equilibration of the system and averaging is done over $3\times
10^5$ MD steps.

\section{Discussion}    \label{sec:discuss}

In a recent experiment, the kinetics of bundle formation of actin in
the presence of multivalent salt has been monitored, using
fluorescence microscopy, and it has been found that the initial
stage of bundle growth tends to saturate when the bundles reach the
size of 10-20 filaments \cite{GH-Lai}. The experiment suggests that
the bundles at this stage very actively exchange single filaments
with the solution, and do not seem to be trapped in a
non-equilibrium state that is hindered by kinetic barriers. It also
shows that at later stages the bundles grow longitudinally, while
still keeping the same thickness \cite{GH-Lai}. A similar kinetic
pattern is observed in our simulations, showing an initial bundle
formation that appears to suddenly ``saturate'' or drastically slow
down in its growth when the size of 10-11 is reached. It should be
noted, however, that in our simulation we did not probe the
equilibrium structure of the system and therefore our results do not
provide conclusive information on the equilibrium distribution of
the bundle sizes. We also note that we did not observe the
longitudinal growth within our simulation time scale, which is
consistent with the experimental observation that the time scale for
the diffusion-limited aggregation (DLA) of the already formed
bundles is 2-3 orders of magnitude longer than the bundle formation
time scale \cite{GH-Lai}. It is worth mentioning, however, that a
Smoluchowski type kinetics does describe even the later stages of
the evolution in the experiment which involves time scale of the
order of {\em hours}, but interestingly this rather long time scale
is indeed mostly set by the diffusion component of the aggregation
time scale $t_p$ [Eq. (\ref{tp-eq})] and the actual energy barrier
to overcome is quite small (of the order of $k_{\rm B} T$)
\cite{GH-Lai}. This suggests that our proposed scheme of quantifying
the kinetics of the growth within the framework of a flocculation
dynamics which takes into account both the availability of the
constituents (controlled by diffusion) and the energetic barriers to
overcome can help us better understand the problem of finite bundle
formation. In other words, just because a part of the process is
slow in absolute time scale it does not necessarily mean that a
large energy barrier is involved. We would like to point out that
this delicate issue has not been recognized in the previous
literature of polyelectrolyte bundles.

One may wonder whether the simple model used here can properly
account for the physics of charged actin protein filaments in
solution. In particular, the diameter of F-actin is one order of
magnitude larger than the value we have used for our model stiff
polyelectrolytes. Looking at the spatial distribution of the charges
on the large protein surface, however, one can note that they are
distributed on narrow twisted strips with a helical pitch that is
considerably larger than the Debye length. This means that the
effective portions of the charge distribution on the actin filaments
that interact with each other are in fact not too different from
thin short rods of the same charge density.
Since the Debye screening length is much smaller than the pitch, one
expects that the twist structure does not matter that much at this
stage. Another important point is that finite-sized bundle formation
has been observed in a variety of bio-polyelectrolytes such as
actin, microtubule etc., each of which having very different
detailed structures \cite{Bundle}. The generality of the observed
phenomenon suggests that the details of the individual systems are
probably not the key determining factor in the formation of finite
bundles. Since all of the bio-polymers that make finite bundles are
highly charged, one is naturally led to the important question of
whether electrostatics alone can cause this effect. This is why
observation of a tendency to form finite bundles in simple model
polyelectrolytes like ours could provide the key to understanding
the physical mechanisms behind this phenomenon.

Finally, we would like to remark on a similar work that has been
recently performed by Sayar and Holm, in which the thermodynamic
properties of condensed bundles with multivalent counterions is
studied \cite{Sayar}. They use a hybrid MC-MD technique, and are
primarily concerned with finding the ultimate equilibrium properties
of the system, rather than the kinetic pathways of going towards
that equilibrium which we probe. In this sense, the two works are
complementary as they approach a similar problem from different
perspectives and with different tools. The main finding of Ref.
\cite{Sayar} is that depending on the parameters the equilibrium
state of such stiff polyelectrolytes could be both finite bundles
and macroscopic condensation. Their results on potential of mean
force can be used to deduce energy barriers, which are in agreement
with our kinetic estimates. There are also slight differences in the
two systems which might cause differences in their behaviors. For
example, in Ref. \cite{Sayar} there are no added salt ions and the
system just has enough counterions to neutralize. This means that
entropic considerations might hamper full neutralization of the
system and thus create additional barriers (for unneutralized
bundles). In our case because we have both salt and counterions, the
monovalent ions can partake the entropy and therefore the
multivalent counterions can happily reside in the bundle areas,
which is perhaps a more faithful modeling of the actual environment
of such systems.

In conclusion, we have studied the kinetics of the bundle formation
for stiff polyelectrolytes in multivalent salt solutions. The
distribution of cluster sizes was found to follow a Smoluchowski
dynamics, with no appreciable energy barrier in the optimal kinetic
paths, in contrast to previous suggestions \cite{HaLiu}. We also
found that the dominant kinetic mode of aggregation comes from
configurations with one end of a rod meeting the other rods at
nearly right angle, and not parallel as proposed in Ref.
\cite{Shkl}. These results could hopefully shed some light on the
controversial issue of the finite size of actin bundles.

\begin{acknowledgments}
It is our pleasure to acknowledge stimulating discussions with
Gerard Wong and assistance from Sarah Mohammadinejad.
\end{acknowledgments}


\end{document}